\documentclass[%
aps,
reprint,
superscriptaddress,
noeprint,
amsmath,amssymb,
prb,
showkeys,
floatfix,nofootinbib
]{revtex4-2}

\usepackage[dvipsnames]{xcolor}
\usepackage{graphicx}
\usepackage[hidelinks=true,colorlinks=true,linkcolor=blue,citecolor=blue]{hyperref}
\usepackage[utf8]{inputenc}
\usepackage[T1]{fontenc}
\usepackage{physics}
\usepackage{mathtools}
\usepackage{siunitx}
\usepackage{bm}
\usepackage{dcolumn}
\usepackage{appendix}
\usepackage{lipsum}
\usepackage{multirow}
\usepackage{placeins}
\usepackage{soul}
\usepackage{pdfpages}
\newcommand{\orcid}[1]{\href{https://orcid.org/#1}{\includegraphics[width=8pt]{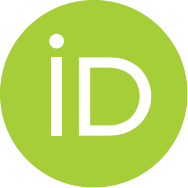}}}
\usepackage[normalem]{ulem}

\makeatletter
\AtBeginDocument{\let\LS@rot\@undefined}
\makeatother

\begin{document}
\title{Linear Jacobi-Legendre expansion of the charge density for machine learning-accelerated electronic structure calculations}

\author{Bruno Focassio\orcid{0000-0003-4811-7729}}\email{b.focassio@ufabc.edu.br}
\affiliation{Federal University of ABC (UFABC), 09210-580 Santo Andr\'e, São Paulo, Brazil}
\affiliation{Ilum School of Science, CNPEM, 13083-970 Campinas, São Paulo, Brazil}
\affiliation{School of Physics and CRANN Institute, Trinity College Dublin, Dublin 2, Ireland}

\author{Michelangelo Domina\orcid{0000-0002-7009-9240}}
\affiliation{School of Physics and CRANN Institute, Trinity College Dublin, Dublin 2, Ireland}

\author{Urvesh Patil\orcid{0000-0002-3924-653X}}
\affiliation{School of Physics and CRANN Institute, Trinity College Dublin, Dublin 2, Ireland}

\author{Adalberto Fazzio\orcid{0000-0001-5384-7676}}
\affiliation{Ilum School of Science, CNPEM, 13083-970 Campinas, São Paulo, Brazil}
\affiliation{Federal University of ABC (UFABC), 09210-580 Santo Andr\'e, São Paulo, Brazil}

\author{Stefano Sanvito\orcid{0000-0002-0291-715X}}\email{sanvitos@tcd.ie}
\affiliation{School of Physics and CRANN Institute, Trinity College Dublin, Dublin 2, Ireland}
\date{\today}

\begin{abstract}
Kohn-Sham density functional theory (KS-DFT) is a powerful method to obtain key materials' properties, but the iterative solution of the KS equations is a numerically intensive task, which limits its application to complex systems. To address this issue, machine learning (ML) models can be used as surrogates to find the ground-state charge density and reduce the computational overheads. We develop a grid-centred structural representation, based on Jacobi and Legendre polynomials combined with a linear regression, to accurately learn the converged DFT charge density. This integrates into a ML pipeline that can return any density-dependent observable, including energy and forces, at the quality of a converged DFT calculation, but at a fraction of the computational cost. Fast scanning of energy landscapes and producing starting densities for the DFT self-consistent cycle are among the applications of our scheme.
\end{abstract}
%
\maketitle

\section*{INTRODUCTION}

As the main workhorse in electronic structure calculations, density functional theory (DFT) \cite{Hohenberg1964,Kohn1965} is today the most widely used method to compute materials properties. Its success derives from the favourable trade-off between computational overheads and accuracy, even when using simple approximations for the exchange and correlation energy functional \cite{Kohn1965,Perdew1992,Perdew1996a,Burke_PerspectiveDFT2012}. The central quantity in DFT is  the electron charge density that, in principle, gives access to the ground-state properties \cite{Hohenberg1964}, and of  particular interest, to the ground-state total energy. In practice, however, the DFT functional is never minimized directly by  using the charge density \cite{Ligneres2005}, but rather by solving a self-consistent set of single-particle equations, known  as the Kohn-Sham (KS) equations \cite{Kohn1965}. This procedure effectively imposes a computational bottleneck and although large-scale calculations can be performed \cite{ScalingCONQUEST,ScalingONETEP}, the typical system routinely 
simulated by DFT rarely reaches a few hundred atoms.

Machine learning (ML) has recently emerged as a surrogate for solving DFT KS equations and possibly replacing them \cite{Bartok2017,Butler2018,Morgan2020}. For instance, trained ML models can be used as predictors for properties such as the energy gaps \cite{Zhuo2018_MLPropPred_bandgap,Rajan2018b_MLPropPred_bandgap,Borlido2020_MLPropPred_bandgap}, superconducting critical temperatures \cite{Stanev2018_MLPropPred_tc,Xie2019_MLPropPred_tc,NelsonSanvito2019_MLCurieTemp,Xie2022_MLPropPred_tc,Zhang2022_MLPropPred_tc}, thermodynamic stability \cite{Bartel2020_MLPropPred_stability}, topological invariant  \cite{Claussen2019_MLPropPred_topo,Focassio2021_MLPropPred_topo}, just to name a few. These models learn a direct map between the structure/composition and the target property, thus avoiding one or many computationally expensive calculations. Using ML for such mapping comes at the cost of accuracy, transferability, physical insight and the need for a large volume of high-quality training data, usually obtained through these very same computationally expensive calculations or, more rarely, from experimental sources \cite{Chibani2020_MLPredProp,Schleder2019_FromDFTtoML}.

For tasks such as structure prediction \cite{Pickard2011_MLStructurePred,Podryabinkin2019_MLStructurePred,Tong2020_MLStructurePred,Wengert2021_MLStructurePred}, phase diagrams evaluation \cite{Deringer2020_MLPhaseDiagram,Muhli2021_MLPhaseDiagram,Gubaev2019_MLMatDiscoveryPhaseDiagram}, molecular dynamics \cite{Rosenbrock2021_MLMatDiscovery,Kresse_MeltingPointMLFF,Reily_MLMDWater}, and, more generally, materials discovery \cite{Jennings2019_MLMatDiscovery,Cobelli2022_MLMatDiscovery,Choudhary2018_MLMatDiscovery,Tkatchenko2020_MLChemDiscovery} one requires fast access to accurate energy, forces and stress tensor of the system investigated. Machine learning inter-atomic potentials (ML-IAPs) are developed to this end, bridging the gap between \textit{ab initio} methods and empirical force fields. The several strategies proposed to date implement a diversity of structural representations and learning algorithms \cite{Behler2007_symmetryfunctions,Behler2011_symmetryfunctions,Bartok2013_SOAP_Bispectrum,Huo2017_MBTR,Bartok2013_SOAP_Bispectrum,Zuo2020_MLFFs,Himanen2020_dscribe} to design ML-IAPs attaining accuracies close to that of DFT at a small fraction of the computational cost \cite{Zuo2020_MLFFs}. The performance of these models is not only a product of the representation of the atomic structure and the ML algorithm, but also the volume, quality, and diversity of the data play a fundamental role \cite{Zuo2020_MLFFs,Unke2021_MLFF}. In general, the construction of ML-IAPs requires campaigns of DFT calculations, whose extension and quality depend on the problem at hand (e.g., the number of species present in a given compound) and the range of applicability of the potential (e.g., the temperature range).

A radically different use of ML consists of improving the theory at its core instead of targeting the DFT outputs. For instance, ML can be used to numerically design new energy density functionals, effectively producing fully exchange and correlation energies \cite{Snyder2012_MLXC,Nagai2020_MLXC,Dick2020_MLXC,Li2021a_MLXC,DeepMind_MLXC_google,NelsonSanvito2019_MLDFTHubbard}, and kinetic energy densities \cite{Snyder2013_MLKEDF,Meyer2020_MLKEDF,Alghadeer2021_MLKEDF,Ryczko2021_MLKEDF}. These strategies, in general, seek to find more accurate approximations to the DFT energy, going beyond the current approximations \cite{Burke2022_MLandDFT_XCfunctionals}, or to eliminate the need of introducing the KS construct by replacing the self-consistent KS equations with a direct minimization of the functional \cite{Ligneres2005}. Unfortunately, although promising, these approaches are still far from obtaining a  ``universal'' functional, treating all systems on an equal footing \cite{Burke2022_MLandDFT_XCfunctionals}. Note also that the construction of ML functionals requires results obtained at the wave-function quantum-chemistry level, a highly computationally expensive task.

In the same spirit, an alternative way to include ML in the DFT workflow is to construct models to directly predict the converged target DFT quantities, namely the Hamiltonian \cite{Zhang2022_Hamiltonian}, the wavefunctions \cite{Schutt2019_MLWFC,Hermann2020_MLWFC} and the electron density \cite{Grisafi2021_SAGPR_transferable_localenvironments,Lewis2021_SALTED,Brockherde2017_burkeDatasetsBenzene,Chandrasekaran2019_ramprasadNNchg_ldos,Ellis2021_AttilaMALA,MLElectronDensity_EquivariantNN}. The goal here is not that of improving the functional, but to reduce or completely eliminate the number of iterative steps needed to solve the KS equations. There are two main approaches used to predict the electronic charge density, $n(\mathbf{r})$,  through ML. One possibility is to expand $n(\mathbf{r})$ over a local-orbital basis set and learn by ML the expansion coefficients. The completeness of such expansion, the basis set details, and the size of the training data limit the accuracy of the ML model \cite{Grisafi2021_SAGPR_transferable_localenvironments,Lewis2021_SALTED} and may introduce errors intrinsic to the particular representation \cite{Brockherde2017_burkeDatasetsBenzene}. Also, the approach is not transferable, namely a different ML model must be constructed for any different basis set. 

The second approach considers the real-space representation of $n(\mathbf{r})$, which is written over a grid in Cartesian space. This is a more ``natural'' representation available in any DFT code. Its main advantage is that the value of the electron density at a grid point is rotationally invariant with respect to the external potential, namely with respect to the position of the surrounding nuclei. As such, one can construct ML models that predict $n(\mathbf{r})$ one grid point at the time, using as descriptors the local atomic neighbourhood of any given grid point (within some chosen cutoff radius). The success of such a grid-based approach largely depends on the chosen representation for the local environment and the learning algorithm. Usually, a single DFT calculation results in tens of millions of grid points so that the generation of abundant training data appears like an easy computational task. However, in a single calculation, there is data redundancy and little diversity (a narrow distribution of external potentials is explored), so multiple configurations for the same systems are usually considered. Then, one typically constructs large neural networks with millions of weights to be learned \cite{Chandrasekaran2019_ramprasadNNchg_ldos,Ellis2021_AttilaMALA}, resulting in generally heavy models with little transferability.

Here our main focus is to transform such a grid-based approach into a lightweight tool that can be universally applied to DFT calculations. This is achieved by drastically reducing the computational overheads while reaching extremely high accuracies. In particular, we introduce a grid-centred representation of the atomic structure based on the Jacobi and Legendre (JL) polynomials, which were previously proposed to construct efficient ML force fields \cite{JLarxiv}. The JL representation is used to build a linear regression for the charge density, where the many-body contributions of different orders are separated. This results in a very compact model with a few coefficients to be trained on a small subset of the total number of grid points available. For the sake of brevity, we call such a class of models Jacobi-Legendre charge density models (JLCDMs). The efficiency and accuracy of our scheme are demonstrated for a range of molecules and solids, including benzene, aluminium, molybdenum, and two-dimensional MoS$_2$. In particular, we show that the KS self-consistent cycle can be bypassed completely in calculating fully converged total energies and forces. Our method is implemented to work with the widely used Vienna \textit{ab initio} simulation package (VASP) \cite{Kresse1996, Kresse1996c}.

\section*{RESULTS}

\begin{figure*}
\centering
\includegraphics[width=\linewidth]{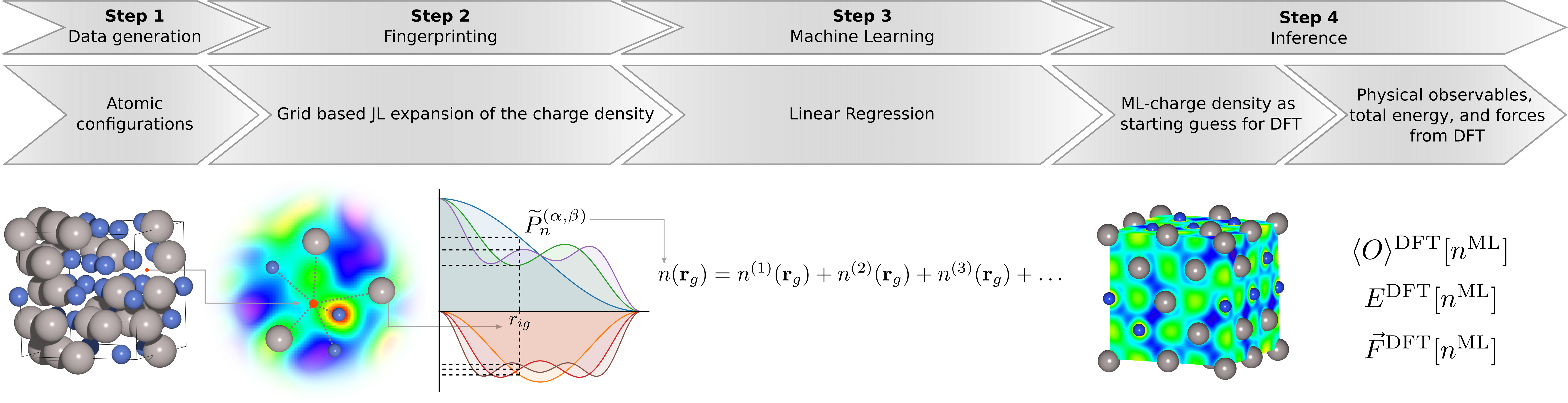}
\caption{Illustration of the workflow used to construct a JLCDM predicting the converged DFT ground-state 
charge density and the associated observables. (Step 1) The procedure starts with an atomic distribution and the mapping of the space over a Cartesian grid. (Step 2) Each grid point is associated with a local atomic environment described by the Jacobi-Legendre expansion. Such expansion is used to construct a linear model (Step 3) that, once trained, accurately predicts the charge density of the grid point. After computing the charge density over the entire grid, this is used to perform DFT calculations (Step 4). For instance, the total energy, the atomic forces, and other density-dependent observables can be easily obtained by using a few steps of frozen-density KS-DFT instead of the full self-consistent cycle.}
\label{fig:workflow}
\end{figure*}

Figure \ref{fig:workflow} provides a schematic view of the construction of a JLCDM. Given an atomic configuration, the space is subdivided into a Cartesian grid, and the atomic environment (the position of the atoms) of each grid is described by an expansion of JL polynomials. A selected number of such expansions forms the training set of a linear regression model that predicts the charge density over the entire grid. Finally, this is used as the converged ground-state density to evaluate energy, forces, and any other density-dependent observable, $\langle O\rangle$, or as a starting point for self-consistent KS-DFT calculations.

\subsection*{Linear expansion of the charge density}

The charge density, $n(\mathbf{r})$, at a grid point $\mathbf{r}_{g}$ can be separated into many-body contributions as

\begin{equation}
    n(\mathbf{r}_{\rm g}) = n^{(1)}(\mathbf{r}_{\rm g}) + n^{(2)}(\mathbf{r}_{\rm g}) + n^{(3)}(\mathbf{r}_{\rm g}) + \ldots + n^{(m)}(\mathbf{r}_{\rm g}) \label{eq:many_body_separation_chg}
\end{equation}

\noindent where $n^{(m)}$ is the $m^{\rm th}$-body ($m$B)  term of the expansion. Thus, $n^{(1)}(\mathbf{r}_{\rm g})$ encodes the atomic contributions to the charge density at $\mathbf{r}_{\rm g}$, $n^{(2)}$ is the contribution from atom pairs, $n^{(3)}$ is the contribution from atoms triplets, etc. Equation \eqref{eq:many_body_separation_chg} can then be rewritten as

\begin{equation}
    n(\mathbf{r}_{\rm g}) = \sum_i n_i^{(1)}(\mathbf{r}_{\rm g}) + \sum_{i\neq j} n_{ij}^{(2)}(\mathbf{r}_{\rm g}) + \sum_{i\neq j, i\neq k, j\neq k} n_{ijk}^{(3)}(\mathbf{r}_{\rm g}) + \ldots 
\end{equation}

\noindent where the sums over the $i$,$j$,$k\ldots$ indexes run over the atoms neighbouring the grid point at $\mathbf{r}_{\rm g}$ up to the cutoff distance, $r_{\rm cut}$. The assumption that the electron density at one point is determined mostly by the external potential generated by the closest atoms follows from the wave mechanics' locality principle \cite{Kohn1978_LocalityPrinciple}.

The atomic configurations required by each contribution in the expansion are expressed through a local representation that here we generally call ``fingerprint''. The fingerprints should be: (i) invariant by translations, (ii) invariant by global rotations of the atoms in the reference frame of the grid point, (iii) invariant to changes in the coordinate system, (iv) invariant to permutations of the atomic indices. Furthermore, they should provide a continuous map of the atomic neighbourhood, i.e., small changes in the atomic structure must reflect small changes in the fingerprints. Finally, the fingerprints should be uniquely determined \cite{CeriottiCsanyi2020_IncompRepresentation} and computationally cheap. 

Following closely reference \cite{JLarxiv}, we expand the one-body contribution, $n_i^{(1)}$, using the distances between the grid point and the atomic neighbourhood as

\begin{equation}
    n_i^{(1)}(\mathbf{r}_{\rm g}) = \sum_{n=1}^{n_{\rm max}} a^{Z_i}_n \widetilde{P}_n^{(\alpha,\beta)}\left( \cos \left( \pi \frac{r_{i{\rm g}} - r_{\rm min}}{r_{\rm cut} - r_{\rm min}} \right)\right) \label{eq:1b_jl}
\end{equation}


\begin{equation}
\widetilde{P}_n^{(\alpha,\beta)}(x) = \left\{\begin{matrix*}[l]
P_n^{(\alpha,\beta)}(x) - P_n^{(\alpha,\beta)}(-1) & \text{for } -1 \leq x \leq 1 \\
0 & \text{for } x < -1 \\
\end{matrix*}\right.
\label{eq:shifted_jacobi_poly}
\end{equation}

\noindent with $P_n^{(\alpha,\beta)}$ being the Jacobi polynomial of order $n$. Here, $r_{i{\rm g}} = |\mathbf{r}_i - \mathbf{r}_{\rm g}|$ is the distance between the grid point $g$ at $\mathbf{r}_{g}$ and the $i$th atom $i$ at $\mathbf{r}_{i}$, $r_{\rm cut}$ is the radius cutoff, $r_{\rm min}$ is a distance shift parameter in the range $(-\infty,r_{\rm cut})$. 
The degree of the expansion is set by $n$ with the sum running in the interval $\left[1,n_{\rm max}\right]$, while $\alpha$ and $\beta$ control the shape of the polynomial with $\alpha,\beta > -1$. 
Note that the choice of the basis used to expand the atomic structure is not unique. Jacobi polynomials represent a convenient one, 
since they effectively describe a vast class of basis-set types. For instance $(\alpha,\beta)=(0,0)$ gives us the Legendre polynomials, while
$(\alpha,\beta)=(0.5, 0.5)$ the Chebyshev polynomials of the second kind. Here we decide to maintain the generality and treat the
$(\alpha,\beta)$ pair as an hyperparameter to optimize. 
The expansion coefficients $a_n^{Z_i}$ in Eq. \eqref{eq:1b_jl} depend on the atomic species considered. As defined in Eq. \eqref{eq:shifted_jacobi_poly}, the ``vanishing Jacobi polynomials'' smoothly vanish at the cutoff radius without needing an additional \textit{ad-hoc} cutoff function.

The terms forming the two-body contribution, $n_{ij}^{(2)}$, can be uniquely written as a function of two distances, $r_{i{\rm g}}$ and $r_{j{\rm g}}$, and the cosine of the subtended angle at $g$, $\hat{\mathbf{r}}_{i{\rm g}} \cdot \hat{\mathbf{r}}_{j{\rm g}}$. We then expand the distances over the vanishing Jacobi polynomials and the angle over Legendre polynomials. The expansion can then be written as,

\begin{equation}
\label{eq:2b_jl}
n_{ij}^{(2)}(\mathbf{r}_{\rm g}) = \sum_{n_1,n_2=1}^{n_{\rm max}} \sum_{l=0}^{l_{\rm max}} a^{Z_iZ_j}_{n_1 n_2 l} \widetilde{P}_{n_1i{\rm g}}^{(\alpha,\beta)} \widetilde{P}_{n_2j{\rm g}}^{(\alpha,\beta)}P_l^{ij{\rm g}},
\end{equation}

\noindent where we have used the shorthand notations $$\widetilde{P}_{ni{\rm g}}^{(\alpha,\beta)} = \widetilde{P}_n^{(\alpha,\beta)}\left(\cos \left( \pi \frac{r_{i{\rm g}} - r_{\rm min}}{r_{\rm cut} - r_{\rm min}} \right)\right),$$ and $P_l^{ij{\rm g}} = P_l(\hat{\mathbf{r}}_{i{\rm g}} \cdot \hat{\mathbf{r}}_{j{\rm g}})$, $P_l$ is the Legendre polynomial, $\hat{\mathbf{r}}_{pg} = (\mathbf{r}_p - \mathbf{r}_{\rm g}) / r_{pg}$, and $l$ defines the Legendre expansion degree with the sum running in the interval $\left[0,l_{\rm max}\right]$. 
The Legendre polynomials are the natural choice for expanding the scalar products between two real space versors in three dimensions, as suggested by the addition theorem of spherical harmonics. As in the one-body case, the expansion coefficients $a^{Z_iZ_j}_{n_1 n_2 l}$ depend on the pair of atomic species considered. The Jacobi indices $n_1$ and $n_2$, and the atom indices $i$ and $j$ are symmetric under the simultaneous swap, therefore if $Z_i=Z_j$ only terms $n_1 \geq n_2$ should be considered. 

Notice that, in the $m$-body expansion for $m>1$, angular information enters via a pairwise dot product of unit vectors joining the atoms to the grid point. The unit vectors are ill-defined when the distance of the grid point from the atom approaches zero and creates a discontinuity in the fingerprints. Assuming that the atomic contribution (1B term) to the charge density dominates at very small distances from the nucleus, we can introduce a double-vanishing Jacobi polynomial in place of the simple vanishing one for all the $m$-body expansions with $m>1$ as given in Eqs. \eqref{eq:2b_jl_doubled_shifted} and \eqref{eq:3b_jl}. The double-vanishing Jacobi polynomials are defined as

\begin{equation}
\overline{P}_{n}^{(\alpha,\beta)}(x) = \widetilde{P}_{n}^{(\alpha,\beta)}(x) - \displaystyle\frac{\widetilde{P}_{n}^{(\alpha,\beta)}(1)}{\widetilde{P}_{1}^{(\alpha,\beta)}(1)}\widetilde{P}_{1}^{(\alpha,\beta)}(x) \; \text{for } n\geq 2
\end{equation}

\noindent with $x = \cos \left( \pi \displaystyle\frac{r_{i{\rm g}} - r_{\rm min}}{r_{\rm cut} - r_{\rm min}}\right)$. 
Equation \eqref{eq:2b_jl} now reads

\begin{equation}
\label{eq:2b_jl_doubled_shifted}
n_{ij}^{(2)}(\mathbf{r}_{\rm g}) = \sum_{n_1,n_2=2}^{n_{\rm max}} \sum_{l=0}^{l_{\rm max}} a^{Z_iZ_j}_{n_1 n_2 l} \overline{P}_{n_1i{\rm g}}^{(\alpha,\beta)} \overline{P}_{n_2j{\rm g}}^{(\alpha,\beta)}P_l^{ij{\rm g}}
\end{equation}

\noindent with $n_1,n_2\geq 2$. Generally, a $m$-body cluster centred on the grid point $g$ can be uniquely defined by $m$ distances and the $m(m-1)/2$ angles subtended at $g$. Using the recipe from Eqs. \eqref{eq:1b_jl} and \eqref{eq:2b_jl_doubled_shifted}, the $m$-body expansion can then be written by associating a Jacobi polynomial to each distance and a Legendre polynomial to each angle. For instance, the three-body contribution $n_{ijk}^{(3)}$ is of the form

\begin{equation}
\label{eq:3b_jl}
\begin{split}
    \MoveEqLeft
    n_{ijk}^{(3)}(\mathbf{r}_{\rm g}) = \sum_{n_1,n_2,n_3=2}^{n_{\rm max}} \sum_{l_1 l_2 l_3}^{l_{\rm max}} a^{Z_iZ_jZ_k}_{n_1 n_2 n_3 l_1 l_2 l_3} \times \\
    &\times \overline{P}_{n_1i{\rm g}}^{(\alpha,\beta)} \overline{P}_{n_2j{\rm g}}^{(\alpha,\beta)} \overline{P}_{n_3k{\rm g}}^{(\alpha,\beta)} P_{l_1}^{ij{\rm g}} P_{l_2}^{ik{\rm g}} P_{l_3}^{jk{\rm g}}
\end{split}
\end{equation}

Using this charge density expansion at each grid point, we can generate a linear representation of the charge density in the expansion coefficients. Therefore, we can learn the ground state charge density by using linear regression, as 



\begin{equation}
\begin{split}
    \MoveEqLeft
    n^{\rm DFT}(\mathbf{r}_{\rm g}) = \sum_i \sum_{n}^{n_{\rm max}} a^{Z_i}_n \; \widetilde{P}_{n_1i{\rm g}}^{(\alpha,\beta)} + \\
    & \sum_{i\neq j} \sum_{n_1 n_2}^{n_{\rm max}} \sum_{l=0}^{l_{\rm max}} a^{Z_iZ_j}_{n_1 n_2 l} \overline{P}_{n_1i{\rm g}}^{(\alpha,\beta)} \overline{P}_{n_2j{\rm g}}^{(\alpha,\beta)}P_l^{ij{\rm g}} + \\
    & \sum_{\substack{{i\neq j,}\\{i\neq k,}\\{j\neq k}}}\sum_{\substack{{n_1 n_2 }\\{n_3}}}^{n_{\rm max}} \sum_{l_1 l_2 l_3}^{l_{\rm max}} a^{Z_iZ_jZ_k}_{\substack{{n_1 n_2 n_3}\\{l_1 l_2 l_3}}} \overline{P}_{n_1i{\rm g}}^{(\alpha,\beta)} \overline{P}_{n_2j{\rm g}}^{(\alpha,\beta)} \overline{P}_{n_3k{\rm g}}^{(\alpha,\beta)} P_{l_1}^{ij{\rm g}} P_{l_2}^{ik{\rm g}} P_{l_3}^{jk{\rm g}} + \\
    & + \ldots
\end{split}
\end{equation}


In the next section, we will demonstrate the prediction power of our JLCDM for a benzene molecule, for periodic solids such as aluminium (Al) and molybdenum (Mo), and a two-dimensional material MoS$_2$. We will also demonstrate the generalisation power of JLCDM for previously unknown phases of Al and MoS$_2$. Finally, we will show that the charge density predicted by our model can be fed back into popular DFT codes to accurately calculate the total energy and forces at a fraction of the typical numerical cost.

\begin{figure*}
\centering
\includegraphics[width=\linewidth]{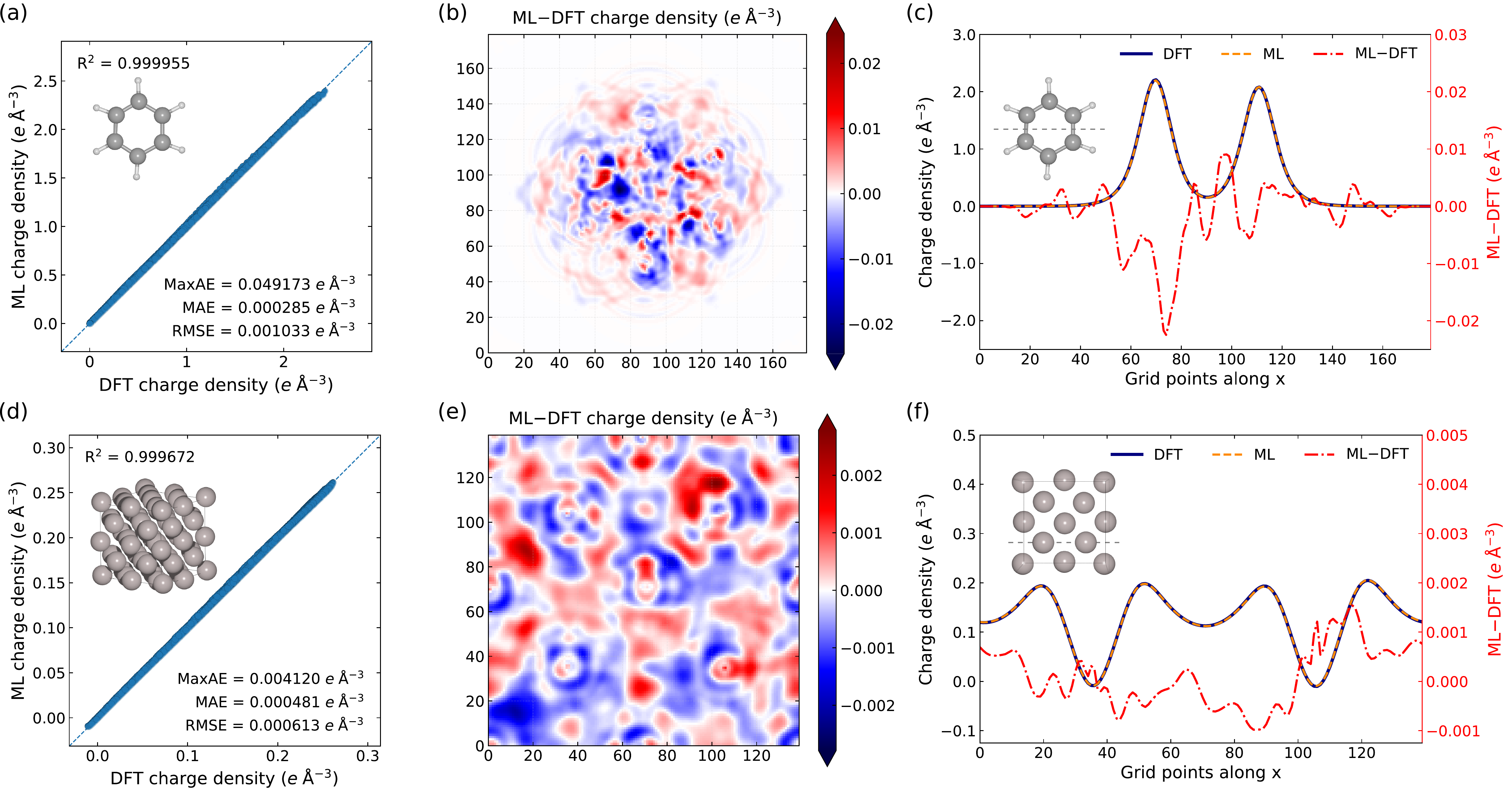}
\caption{Analysis of the performance of the JLCDM. Panel (a) displays the parity plot for the benzene test set together with mean absolute error (MAE), root-mean-squared error (RMSE), maximum absolute error (MaxAE) and $R^2$ metrics. Panel (b) displays the charge density difference (the error) between the fully converged DFT ground-state density and that predicted by the model for a distorted benzene configuration selected from the test set. The results for a symmetric benzene molecule can be found in the Supplemental Information (SI). Here we show the plane containing the molecule. Panel (c) shows DFT and JLCDM-predicted charge density for benzene computed along the line indicated in the inset. The plot also reports their difference with values provided on the right-hand side scale (red). Panel (d) displays the parity plot for the aluminium test set. Panel (e) displays the charge density difference (error) between the fully converged DFT ground-state density and that predicted by the model for a distorted aluminium configuration selected from the test set. The slice shows the basal plane of the supercell ($z=\SI{0.0}{\angstrom}$). In (f) the DFT and ML charge density for aluminium is computed along the line indicated in the inset. The plot also reports their difference with values provided on the right-hand side scale (red). The planes chosen in panels (c) and (f) are the same as those in (b) and (e), respectively.}
\label{fig:parity_diff_al_benzene}
\label{fig:line_diff_al_benzene}
\end{figure*}

\subsection*{Grid-point sampling strategy}

We start our analysis by discussing the construction of an appropriate training set for our JLCDM, which is truncated at the $2$-body order since this is already enough for extremely accurate predictions. Previously published works \cite{Chandrasekaran2019_ramprasadNNchg_ldos,Ellis2021_AttilaMALA} have trained large neural networks over the entire grid-point mesh, typically containing a few million density values. Here we show that this is not necessary since there is significant redundancy in the information, and often the inclusion of the entire density in the training set has just the effect of producing an unbalanced ensemble. This is easy to see in the case of molecules, where most grid points are situated far away from the molecule and, by sitting in a vacuum, possess similar vanishing small charge density. For this reason, we implement a sampling strategy that allows us to use only a small fraction of the grid points but includes more diverse atomic arrangements.

In practice, our simple sampling scheme consists in assigning to a point $\mathbf{r}$ in space a probability of selection based on the value of the charge density, $n(\mathbf{r})$, at that point. The probability of selection is given by a normal distribution of the inverse of the charge density, namely $\exp[-(1/n(\mathbf{r}))^2/2\sigma^2]$. This choice gives more importance to grid points presenting large electron densities, while low-density regions will contribute little to the training set. 
The parameter $\sigma$ controls how sharp or broad this probability distribution is, a tool that helps us to select grid points closer or farther away from the charge density maxima. Such a targeted sampling technique is accompanied by uniform sampling across the unit cell, which guarantees that enough diversity is maintained in the training set. As a result, we can construct an accurate model trained with just about 0.1\% of the available training points (see the Methods section for more details). 
Note that our sampling strategy is not limited to linear charge density expansions. The same can be used as an efficient way to train even neural network models, resulting in much smaller models attaining the same or higher accuracy.

\subsection*{Accuracy of the models}

We now discuss the accuracy that can be reached by the JLCDMs for both molecules and solids. Figure~\ref{fig:parity_diff_al_benzene}(a) displays the parity plot of the charge density at the grid points for the 30 atomic configurations contained in the test set of the benzene molecule. These have been obtained from Ref. \cite{Brockherde2017_burkeDatasetsBenzene} by molecular dynamics at $300$~K and performing DFT calculations on each sampled geometry. For benzene, our 1B$+$2B JLCDM contains 1,572 coefficients trained over 6,000 density-grid points, out of the 5,832,000 available per atomic configuration over the 30 configurations used for training and another 30 for testing. The test-set mean absolute error (MAE) achieved is $0.000285\;e\si{\per\cubic\angstrom}$. Such error corresponds to $0.011\%$ of the maximum density, meaning that the charge density obtained by the JLCDM is very close to that of a well-converged DFT calculation. Note that the MAE on the total electron count is $0.025$. The model and sampling hyperparameters are reported in Table \ref{tab:model_hyperparameters} and Table \ref{tab:model_sampling_params}, respectively, in the Methods section.

In panel (b) of Fig. \ref{fig:parity_diff_al_benzene}, we present the difference between the charge density obtained with JLCDM and the converged DFT charge density on a plane, while panel (c) shows a line scan in the same plane of the two charge densities and their difference. As expected, the absolute error is larger in the region closer to the nuclei, where the charge density is maximised. However, no emerging pattern indicates that the JLCDM is not biased against any particular local atomic configuration. Importantly, the error, as the density, vanishes for positions far from the molecule. Our constructed JLCDM performs better than published models \cite{Chandrasekaran2019_ramprasadNNchg_ldos} despite being trained over a tiny fraction of data and being constructed on only 1,572 trainable parameters.

Next, we move to metallic solids, aluminium and molybdenum. Aluminum is a benchmark system chosen for comparison with previously published models \cite{Lewis2021_SALTED,Chandrasekaran2019_ramprasadNNchg_ldos,Ellis2021_AttilaMALA}. Its electronic structure features a very delocalized charge density, so as a second example, we also consider an early transition metal, Mo, which presents a higher degree of charge localization. In constructing the JLCDM, we use the same density sampling procedure as used for the benzene molecule. See the Methods section for details.

In the case of Al, we train and test over 10 configurations obtained from \textit{ab initio} molecular dynamics (AIMD). We find that a 1B$+$2B JLCDM with only 120 trainable coefficients gives us a MAE of $0.000481\;e\si{\per\cubic\angstrom}$, at par with previously published deep neural networks \cite{Chandrasekaran2019_ramprasadNNchg_ldos,Ellis2021_AttilaMALA}. Most importantly, our model generalises better, as we will show in the next section. Figure \ref{fig:parity_diff_al_benzene}(d) shows the parity plot for the Al test set, demonstrating the accuracy achieved. Similarly to the case of benzene, the difference between the ML predictions and the DFT charge density does not present any clear error pattern, see Fig. \ref{fig:parity_diff_al_benzene}(e), except for the expected increase close to the nuclei. In general, the charge density error for Al is found to be 10 times smaller than that found for benzene, as one can see from the line plot of Fig. \ref{fig:line_diff_al_benzene}(f).

Similar results are also obtained for Mo, where a JLCDM with 812 trainable parameters returns a MAE and a RMSE of $0.001974\;e\si{\per\cubic\angstrom}$ and $0.002820\;e\si{\per\cubic\angstrom}$, respectively, see the Supplemental Information (SI) for details. In contrast to benzene and aluminium, the charge density error appears to have a radial distribution centred around each atom with a minimum error in the interstitial region. The maximum absolute error over the test set in this case is only ${\sim}0.06\;e\si{\per\cubic\angstrom}$, and it is found over a small set of grid points.

\begin{figure*}
\centering
\includegraphics[width=\linewidth]{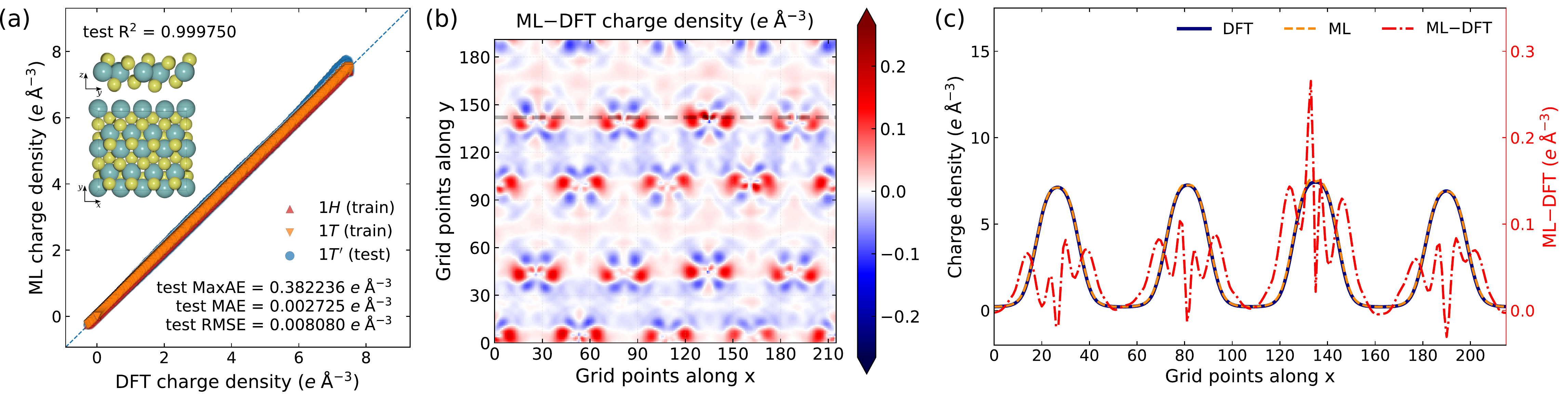}
\caption{Analysis of the performance of the JLCDM for MoS$_2$. Panel (a) shows the parity plot between JLCDM-predicted and DFT charge density for the three MoS$_2$ polymorphs, 1H, 1T, and 1T$^\prime$. In this case 1H and 1T phases are used for training, while the model is tested on the 1T$^\prime$. All the error metrics shown, $R^2$, MAE, RMSE, and MaxAE, correspond to the test set. The inset depicts a snapshot of 1T$^\prime$-MoS$_2$. Panel (b) display the charge density difference (error) between the fully converged DFT charge density and the JLCDM-predicted one over the plane of the monolayer ($z=c/2\;\si{\angstrom}$) for a distorted 1T$^\prime$-MoS$_2$ configuration selected from the test set. Panel (c) shows the charge density profile for fully converged DFT and JLCDM-predicted charge density along the path highlighted with a dashed line in panel (b). The difference between charge densities 
can be read on the right-hand side scale (red) of panel (c).}
\label{fig:mos2_parity_map_line}
\end{figure*}

Finally, we focus on two-dimensional MoS$_2$, which helps us to demonstrate the capability of our JLCDM to generalise to previously unseen phases. Two-dimensional MoS$_2$ can be found in multiple polymorphs, both semiconducting and metallic. Also, for MoS$_2$, we use the same charge-density sampling procedure adopted for benzene, Al and Mo. See the Methods section. However, this time we train and test the model on different phases; namely, the training set is constructed using atomic configurations of the 1H and 1T phases while we test our prediction on the $1\text{T}^\prime$ phase. The 1H phase is formed by sandwiched hexagonal layers of S--Mo--S in a Bernal stacking, while the 1T phase presents a rhombohedral arrangement \cite{Eda2012_mos2_exp}.  As the free-standing 1T phase is unstable, a spontaneous lattice distortion in the $x$ direction creates the $1\text{T}^\prime$ one \cite{1Tmos2_metastable,Eda2012_mos2_exp}, which is depicted in the inset of Figure \ref{fig:mos2_parity_map_line}(a). The three polymorphs present completely different electronic structures. The 1H phase is semiconducting with a \SI{1.58}{\electronvolt} theoretical energy gap, while the 1T phase is metallic \cite{band_structure_tmds_phases}. In contrast, the $1\text{T}^\prime$ polymorph has a topological gap (\SI{0.08}{\electronvolt}) induced by spin-orbit coupling \cite{Xiaofeng2014a_QSHI}, while it remains a semi-metal in the absence of spin-orbit coupling interaction. 

In order to train our JLCDM, we use 10 AIMD (at 300 K) configurations each for the 1H and 1T phases, while the test set is made of ten 1T$^\prime$ AIMD (at 300 K) snapshots. Figure \ref{fig:mos2_parity_map_line}(a) shows the parity plot for all three polymorphs, namely for the training and test set. By visual inspection, one can notice that the error slightly increases for the $1\text{T}^\prime$ phase, but the JLCDM still performs extremely well, displaying a MAE and a RMSE of $0.002725\;e\si{\per\cubic\angstrom}$ and $0.008080\;e\si{\per\cubic\angstrom}$, respectively. Also, the JCDM remains compact with 2,346 trainable parameters in this case. The charge density difference plot, see Fig. \ref{fig:mos2_parity_map_line}(b), tells us that the error tends to be larger in the region around the Mo ions pointing towards the S atoms. This feature is somehow expected since the bonding structure of the three phases is different, trigonal prismatic for 1H, octahedral for 1T phase, and a distorted lattice for $1\text{T}^\prime$. The line density plot of Fig. \ref{fig:mos2_parity_map_line}(c) further shows that the JLCDM slightly overestimates the charge density surrounding the Mo atoms. However, it is worth noting that the error is small, $<2\%$, so the JLCDM-predicted charge density for the unseen $1\text{T}^\prime$ phase is still of high quality, namely, the JLCDM can be used to explore new phases.

\subsection*{JLCDM performance on the DFT total energy and forces}

In the previous section, we have shown that the charge density predicted by our JLCDM is close to the DFT converged one. Now we show that the energy and forces corresponding to such charge density are close to the corresponding converged values, with the average error matching those of state-of-the-art machine-learning force fields.

This is demonstrated by constructing the KS Hamiltonian corresponding to the JLCDM-predicted charge density. The band energy contribution to the total energy, $E_{\rm band} = \sum_i f(\epsilon_i)\epsilon_i$, is obtained by summing up the occupied KS eigenvalues, $\epsilon_i$ [$f(\epsilon_i)$ is the occupation number], which are computed by diagonalizing the KS Hamiltonian. The remaining contributions to the total energy are obtained directly from the JLCDM electron density. Such a scheme is implemented in the VASP code, where an interactive matrix-diagonalization procedure requires performing a set of non-self-consistent iterations to compute the KS eigenvalues and eigenvectors, i.e. the charge density is not updated during these iterations. In this work, we select 5 non-self consistent iterations for all systems. The total energy and forces obtained are then compared with those computed through a converged fully self-consistent DFT calculation. As given by such procedure, the total energy yielded by the JLCDM-predicted charge density may be lower than the KS-DFT ground-state energy.

\begin{table}
\centering
\caption{JLCDM performance metrics on the task of predicting total energy and forces. These are obtained through non-self-consistent DFT using the JLCDM-predicted charge density. The force error is computed over all and all atoms.}
\label{tab:model_performance}
\bgroup
\def\arraystretch{1.25}
\begin{tabular}{lccccc}
\hline
\multicolumn{1}{c}{System} & \multicolumn{2}{c}{Total energy} &  & \multicolumn{2}{c}{Forces} \\
 & \multicolumn{2}{c}{(meV per atom)} &  & \multicolumn{2}{c}{(eV \AA$^{-1}$)} \\ \cline{2-3} \cline{5-6} 
        & MAE   & RMSE   &  & MAE   & RMSE  \\ \hline
Benzene & 4.021 & 4.065  &  & 0.031 & 0.046 \\
Al      & 0.046 & 0.054  &  & 0.007 & 0.009 \\
Mo      & 0.203 & 0.212 &  & 0.019 & 0.024 \\
$1T^\prime$-MoS$_2$ & 8.058 & 8.845  &  & 0.078 & 0.104 \\ \hline
\end{tabular}
\egroup
\end{table}

\begin{figure}
\centering
\includegraphics[width=\linewidth]{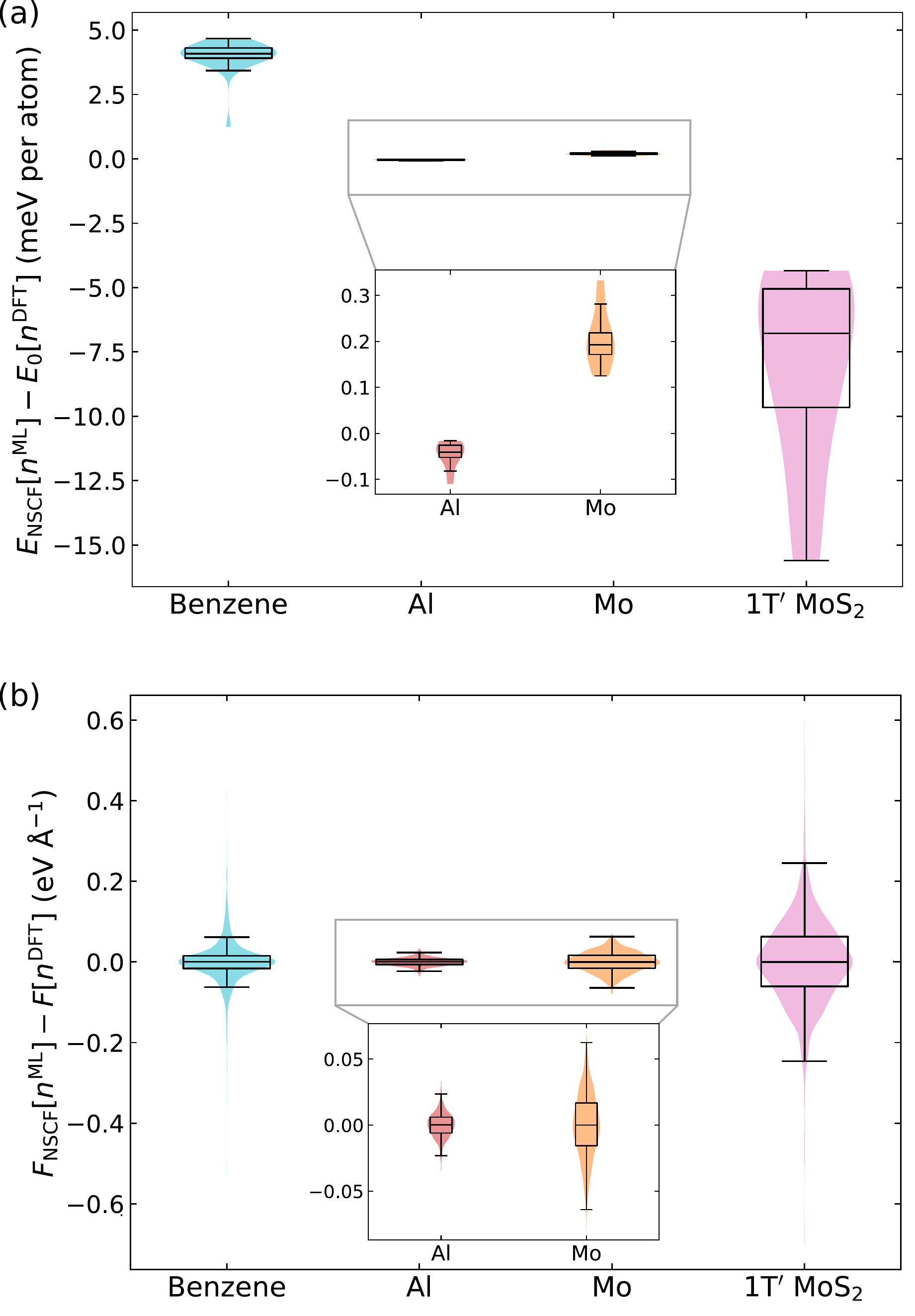}
\caption{Error on the total energy and forces. Box and violin plots for the error on the total energy (a) and the forces (b) computed from JLCDM-predicted charge density. The fully converged DFT values provide the ground truth. The insets show a magnified version of the results for Al and Mo, whose distributions are very narrow on the global scale. The associated absolute mean values are reported in Table \ref{tab:model_performance}. The lines in the middle of the boxes mark the medians. The boxes are plotted from the first to the third quartile, with the line marking the median. The whiskers extend to $1.5$ times the box length.}
\label{fig:error_energy_forces}
\end{figure}

The MAE and RMSE metrics of the calculated energy and forces are given in Table \ref{tab:model_performance}, while Fig. \ref{fig:error_energy_forces} shows the error distributions as box and violin plots. Aluminium presents the narrower total-energy error spread, with values ranging from $-0.11$ meV per atom to $-0.02$ meV per atom and with a mean error at $-0.05$ meV per atom. This is then followed by Mo, with a total-energy error spread between $0.12$ meV per atom and $0.33$ meV per atom with a mean error at $0.20$ meV per atom, and then benzene, with a total-energy error between $1.24$ meV per atom and $4.67$ meV per atom with mean error at $4.02$ meV per atom. Finally, the unseen 1T$^\prime$ phase of MoS$_2$ returns an error range of $-15.60$ meV per atom to $-4.34$ meV per atom and a mean error of $-8.06$ meV per atom. These errors are all very competitive with that achieved by linear ML force fields constructed with a comparable range of parameters \cite{LunghiSanvito2019_CovalentBondFF}.

Next, we investigate the ability of our JLCDM to perform over systems never seen before. Our test is constructed for Al, for which we were able to build the best model, and consists in computing the total energy and forces of a series of 256-atom supercells taken from Ref. \cite{Ellis2021_AttilaMALA}. This dataset contains 10 configurations corresponding to solid Al at 298 K and 10 configurations of both solid and liquid Al at its melting temperature of 933 K. The JLCDM used here is the same discussed before that produced the results from Figure \ref{fig:line_diff_al_benzene}(d)-(f), trained over 32-atom supercells for solid Al at 300 K. Table \ref{tab:model_scaling} summarizes our results. The error on the total energy and forces slightly increases when considering systems in the same conditions but different cell sizes, namely comparing the 32-atom and the 256-atom supercells for solid Al at 300 K and 298 K, respectively. In any case, the MAE remains below 1 meV per atom for the total energy and below 0.025 eV \AA$^{-1}$\ for the forces. As the structures tested become increasingly different from those used for training (data at 933 K) the error grows further, reaching 35.062 meV per atom and 0.164 eV \AA$^{-1}$\ in the liquid phase.

\begin{table}[!ht]
\centering
\caption{Performance of the JLCDM for Al, trained over 32-atom supercells at room temperature, against 256-atom supercells at various conditions. The configurations for the 256-atom supercells are taken from Ref. \cite{Ellis2021_AttilaMALA,mala_data_rodare}, and the test error is computed over 10 samples for each different condition.}
\label{tab:model_scaling}
\bgroup
\def\arraystretch{1.25}
\begin{tabular}{clccccc}
\hline
\multirow{2}{*}{\# atoms} & \multicolumn{1}{c}{\multirow{2}{*}{Condition}} & \multicolumn{2}{c}{Total energy} &  & \multicolumn{2}{c}{Forces} \\
 & & \multicolumn{2}{c}{(meV per atom)} &  & \multicolumn{2}{c}{(eV \AA$^{-1}$)} \\ \cline{3-4} \cline{6-7} 
       & & MAE   & RMSE   &  & MAE   & RMSE  \\ \hline
32  & solid (300 K) & 0.046 & 0.054  &  & 0.007 & 0.009 \\
256 & solid (298 K) & 0.843 & 0.908 &  & 0.025 & 0.031 \\
256 & solid (933 K) & 6.976 & 7.526 &  & 0.068 & 0.862 \\
256 & liquid (933 K) & 35.062 & 36.498 &  & 0.164 & 0.203 \\ \hline
\end{tabular}
\egroup
\end{table}

In order to put our results in perspective, neural network models (${\sim}10^6\text{--}10^7$ trainable weights) using the bispectrum components to describe the local environments reach a MAE of 123.29 meV per atom over the liquid phase, when trained on high-temperature solid structures only \cite{Ellis2021_AttilaMALA}. This means that, on the same test, our JLCDM outperforms neural networks by a factor of four, despite consisting of only 120 trainable parameters and being trained on the 0.1\% of the charge density points. The neural network error is then reduced to 13.04 meV per atom only when the training is performed on both high-temperature solids and liquids \cite{Ellis2021_AttilaMALA}. Certainly, we could systematically improve the JLCDM by adding more distorted supercells in our training set or by including both solid and liquid configurations at 933 K. However, here, we wish to point out that the smooth description of the local environment allows us to achieve very competitive accuracy (35 meV per atom for liquid Al at 933 K) even for such a compact model.

\section*{Discussion}

Inspired by the recently developed Jacobi-Legendre potentials \cite{JLarxiv}, we have designed a grid-based many-body linear expansion of the charge density, where the local external potential is described by Jacobi and Legendre polynomials. The method, combined with a charge-density targeted sampling strategy, produces highly accurate charge densities despite being constructed over an extremely limited number of trainable coefficients. 
We have demonstrated the efficacy of the JLCDM for diverse examples, namely a benzene molecule, solid and liquid Al, solid Mo and different phases of 2D MoS$_2$. In all cases, simple two-body JLCDMs accurately predict the charge density and can be transferred to different phases not originally included in the training set. For instance, training over the 1H and 1T phases of 2D MoS$_2$ is enough to predict the charge density of the 1T$^\prime$ phase, and so is the case for liquid Al, whose density can be constructed from a model trained over solid-state configurations at room temperature. The JLCDM-predicted densities can then be used to compute total energy and forces, achieving accuracy comparable to state-of-the-art machine learning force fields and, in some cases, even to fully converged DFT calculations. 

As it stands, the methodology introduced here could be readily used in a diverse set of applications. If one is interested solely in energies and forces, learning the charge density probably will not be the optimal way to address the problem because of the computational overheads involved in many of the steps required for training and predicting the charge density over a fine grid. In that case, ML force fields can be a better solution, even though the numerical effort to generate the training set needed to achieve DFT accuracy is typically rather extensive, much larger than that required to generate a JLCDM. In any case, a JLCDM strategy becomes essential when one targets density-related electronic quantities, which can be obtained only by DFT. For instance, one may need to evaluate the dipole moment (the Bader charges, the polarizability, etc.) along a set of molecular dynamics trajectories. In that case, a successful strategy may be to use a force field to generate the trajectories and the JLCDM method to evaluate the charge density and the associated properties. Furthermore, applications such as crystal structure prediction, phase diagram construction, reaction path search, and other computationally intensive tasks could be greatly accelerated by using JLCDM-predicted charge densities as the starting point of DFT calculations. Finally, the predicted charge density can be easily employed as the starting density for computationally expensive hybrid-functional calculations.

\FloatBarrier
\section*{METHODS}\label{sec:methods}

\subsection*{DFT calculations and dataset generation}
All single-point and \textit{ab initio} molecular dynamics (AIMD) calculations are performed using density functional theory (DFT) \cite{Hohenberg1964, Kohn1965} as implemented in the Vienna \textit{ab initio} simulation package (VASP) \cite{Kresse1996, Kresse1996c}. Exchange and correlation interactions are treated by the generalized gradient approximation (GGA) \cite{Perdew1992} with the Perdew-Burke-Ernzerhof (PBE) \cite{Perdew1996a} exchange and correlation functional. We use the projector augmented wave (PAW) \cite{Kresse1999a} pseudopotentials. Single-point self-consistent calculations are performed with a \SI{600}{\electronvolt} kinetic-energy cutoff for the plane-wave expansion, and the Brillouin zone is sampled over a $k$-point density of $12\;/$\si{\per\angstrom}. AIMD runs are performed with a \SI{2}{\femto\second} time-step, and the Nosé-Hoover thermostat \cite{Nose1984, Nose1984b, Hoover1985} maintains the $NVT$ ensemble. All AIMD runs are at least \SI{4}{\pico\second} long, and snapshots are taken from the simulation's last \SI{3}{\pico\second}. For benzene and 2D MoS$_2$ sufficient vacuum space, at least \SI{15}{\angstrom} is included in the non-periodic directions so to avoid spurious interaction between periodic images.

\paragraph*{Benzene data generation.}
Data for benzene are extracted from the dataset available at \url{http://quantum-machine.org/datasets/} \cite{Brockherde2017_burkeDatasetsBenzene}. For the training set, we randomly select 30 snapshots from a MD run at 300 K and 400 K, available in the ``\verb|benzene_300K-400K.tar.gz|'' file, and for the test set, 30 snapshots are randomly sampled from MD at 300 K, available in ``\verb|benzene_300K-test.tar.gz|''. The charge density for the selected snapshots is then calculated using VASP with the settings described above. Using \SI{600}{\electronvolt} as the kinetic-energy cutoff for the plane-wave expansion, this results in the charge density being represented over a $180\times 180\times 180$ grid (5,832,000 grid points).

\paragraph*{Al, Mo, and 2D MoS$_2$ data generation.}
For Al, Mo and MoS$_2$, we randomly extract snapshots from AIMD runs at 300 K. For Al, we use a $2\times 2\times 2$ conventional \textit{fcc} supercell containing 32 atoms, while a $3\times 3\times 3$ conventional \textit{bcc} supercell containing 54 atoms described Mo. A $3\times 3\times 1$ supercell is used for the 1H and 1T phases of MoS$_2$ (27 atoms), while for the $1T^\prime$, we consider a $4\times 2 \times 1$ supercell (48 atoms). For Al and Mo, we extract 10 snapshots for training and 10 for testing. For MoS$_2$, we extract 10 snapshots for each phase, with the 1H and 1T structures used for training and 1T$^\prime$ for testing. The charge densities are represented over a $140\times 140 \times 140$ grid (2,744,000 grid points) for Al, a $160\times 160\times 160$ grid (4,096,000 grid points) for Mo, $160\times 160\times 300$ 
 grid (7,680,000 grid points) for MoS$_2$ 1H and 1T, and $216\times 192\times 300$ grid (12,441,600 grid points) for 1T$^\prime$-Mos$_2$.

In order to investigate the transferability of the JLCDM for Al, we use the snapshots reported in Ref. \cite{Ellis2021_AttilaMALA}, as available in \cite{mala_data_rodare}. These Al are 256-atom Al supercells whose charge density has been recalculated with VASP. The energy cutoff for those is lowered to \SI{360}{\electronvolt} so as to match the same real-space grid used in ref. \cite{Ellis2021_AttilaMALA}, $200\times 200 \times 200$ (8,000,000 grid points), and only the $\Gamma$-point is used to sample the BZ.

\subsection*{DFT calculations with fixed charge density}

In order to use the ML charge density to compute total energies and forces, we use KS-DFT while keeping the charge density fixed and using the same settings as specified for the data generation. The ML charge density is kept constant at each step of an iterative diagonalization of the Kohn-Sham Hamiltonian. In particular, the Kohn-Sham eigenstates and eigenvalues are optimized during five steps with no updates to the charge density. A comparison for the Al test set at each step of the (non-) self-consistent cycle can be found in the SI.

While using PAW pseudopotentials, one is required to provide the augmentation on-site occupancies at the start of a calculation. For Al, we ignore one-centre correction terms evaluated on the radial support grid, a strategy that allows us to use the charge-density predictions for unknown structures or arbitrary sizes. For the other systems, we reuse the already known one-centre occupation DFT-computed terms together with our ML charge density to start the new calculations for configurations on the test set. In the future, the augmentation occupancies can also be learned with a similar scheme as designed here. This will allow the use of the ML charge density as a starting point for DFT calculations of any structure.

\subsection*{Model training, hyperparameter optimization and timing}

We fit the linear models by using singular value decomposition to find the pseudo-inverse of $A$ solving the matrix equation, $A\hat{x}=\hat{b}$, for the coefficients $\hat{x}$. Training and inference are performed using the Ridge class (with $\alpha=0$) from the scikit-learn library \cite{scikit-learn}.

Hyperparameter optimization is performed through Bayesian optimization using Gaussian Processes (\verb|gp_minimize|), as implemented in the scikit-optimize library \cite{scikit-opt}. This is done solely on part of the training set. For the Al and Mo JLCMDs, 8 training snapshots are used for training and the remaining 2 are for validation. For benzene, 27 are used for training, and 3 for validation. On MoS$_2$, we take one training snapshot for each phase (1H and 1T) as the validation set, and the remaining training snapshots are used for training. The optimization targets the minimization of the mean absolute error (MAE). Table \ref{tab:model_hyperparameters} shows the hyperparameters for each model. 

\begin{table}[!h]
\centering
\caption{Optimized hyperparameters and corresponding feature size for each model generated.}
\label{tab:model_hyperparameters}
\bgroup
\def\arraystretch{1.25}
\begin{tabular}{lcccccccc}
\hline
\multicolumn{1}{c}{System} & Body  & $r_{\rm cut}$ & $n_{\rm max}$ & $l_{\rm max}$ & $r_{\rm min}$ & $\alpha$ & $\beta$ & \# features \\ \hline
Benzene & 1b & 2.80 & 27 & -  & $-0.78$ & 7.00    & 0.00 & \multirow{2}{*}{1572} \\
        & 2b & 2.80 & 12 & 5  & $ 0.00$ & 7.00    & 0.00 &  \\
Al      & 1b & 4.08 & 15 & -  & $-0.74$ & 7.87    & 3.62 & \multirow{2}{*}{120} \\
        & 2b & 4.08 & 6  & 6  & $ 0.00$ & 5.87    & 1.75 &  \\
Mo      & 1b & 4.04 & 20 & -  & $-1.09$ & 4.02    & 5.46 & \multirow{2}{*}{812} \\
        & 2b & 4.04 & 12 & 11 & $ 0.00$ & $-0.08$ & 2.38 &  \\
2D MoS$_2$ & 1b & 4.76 & 18 & -  & $-0.93$ & 6.72 & 6.97 & \multirow{2}{*}{2346} \\
                    & 2b & 4.76 & 11 & 10 & $ 0.00$ & 5.07 & 2.69 & \\ \hline
\end{tabular}
\egroup
\end{table}

\begin{table*}[!htb]
\centering
\caption{Timing assessment of the different steps needed to train the JLCDM and to perform new predictions with it. The time taken to create the training and test data (selecting the grid points using smart sampling and creating the fingerprints) refers to the 10 snapshots used for each data set partition. The fingerprint calculation is performed in the serial model (single CPU), while DFT calculations are performed with an MPI-parallel version of the VASP code run on 16 cores with the same parallelisation settings. Note that the calculation for the 256-atom Al cell was performed with a JLCDM trained on the 32-atom cell so that there is only one training time.}
\label{tab:timing}
\begin{tabular}{llcc}
\hline \hline
& & Time (s) & Time (s) \\
& & 32-atom Al cell & 256-atom Al cell \\
& & ({2,744,000} grid points) & ({8,000,000} grid points) \\ \hline
\multicolumn{2}{l}{Training} & & \\
 & Sample grid points with smart sampling (training \& test) & 9.401 & -- \\
 & Generate fingerprint vectors from atom positions (training \& test) & 25.424 & -- \\
 & Fitting linear model & 3.449 & -- \\
 & {Total training time} & {38.274} & -- \\
 &  &  &  \\ 
\multicolumn{2}{l}{Inference} & & \\
 & Generate fingerprint vectors from atoms' positions & 247.105 & 710.196 \\
 & Construct density using JLCDM & 0.085 & 0.088 \\
 & DFT calculation using the JLCDM-predicted charge density & 365.660 & 765.928 \\
 & {Total inference time} & {612.765} & {1,476.124} \\ 
 &  &  &  \\ 
 \multicolumn{2}{l}{Fully converged SCF calculation} & {884.930} &  {2,579.836} \\
&  &  &  \\ 
 \multicolumn{2}{l}{JLCDM saving} & 272.080 & 1,103.624 \\
 & & (30.75\%) & (42.78\%) \\ \hline \hline
\end{tabular}
\end{table*}

An assessment of the time needed to train our model and to perform new predictions is provided in Table~\ref{tab:timing}, where this is compared to the time needed to perform a fully converged self-consistent (SCF) calculation for the same system. This enables us to evaluate the time saving achieved over a single non-self-consistent calculation performed with our JLCDM-predicted charge density. Note that an estimate of the total computation time saving, which is inclusive of the effort needed to generate the training set, strongly depends on the specific workflow one aims to follow since it scales with the total number of calculations to perform. Besides the DFT calculations, computing the JL fingerprints dominates the overall inference time. Importantly, we notice that the time taken for the inference of both a 32-atom and a 256-atom supercell is always inferior to the time of the full SCF-DFT calculation. As the size of the target system for inference increases, one see a much more pronounced difference between the time needed for a JLCDM calculation and that for a full SCF-DFT one.

\subsection*{Grid sampling}

The grid points included in the training set are selected by randomly sampling the real-space charge density according to a combination of uniform sampling and targeted sampling on the grid. Targeted sampling is performed by assigning to each grid point, $\mathbf{r}_{\rm g}$, a probability $P$, given by a normal distribution of the inverse of the charge density at that grid point:

\begin{equation}
    P(\mathbf{r}_{\rm g}) = \frac{1}{\sqrt{2\pi}\sigma} \exp \left(-\frac{(1/n(\mathbf{r}_{\rm g}))^2}{2 \sigma^2}\right) \label{eqn:prob_targeted_sampling}
\end{equation}

Targeted sampling is combined with uniform sampling across the simulation cell, composing the training data for each snapshot. The number of grid points sampled through targeted and uniform sampling is manually tuned to better sample the features of each example's charge density. The hyperparameter $\sigma$ is also tuned manually for each example. However, these hyperparameters can readily be included in the automated hyperparameter optimization routine, making it easy to address any molecule or solid-state system. Table \ref{tab:model_sampling_params} shows the parameters used for sampling and the percentage of the available grid point used to train the models. As shown in the Results section, our model requires a very modest data set size compared to other grid-based approaches in the literature while attaining accurate predictions.

In the SI we show the learning curve of the JLCD model for benzene with respect to the number of grid points, comparing our sampling technique with uniform sampling. Models trained with targeted-sampling data exhibit lower errors compared to those trained with uniform sampling, therefore, model accuracy is enhanced by using targeted sampling. In addition, the maximum absolute error for uniform sampling is significantly higher than that of targeted sampling, suggesting that the uniform sampling model has regions in space with poor prediction accuracy. We expect that as the data volume increases, the difference between uniform and targeted sampling diminishes. Nonetheless, we emphasise that our ability to use targeted sampling to develop highly accurate models with minimal data contributes to the efficiency of our workflow.

\begin{table}[!h]
\centering
\caption{Sampling hyperparameters and percentage of used data from the total data available. The percentage of uniform sampling is retrieved out of the percentage of used data.}
\label{tab:model_sampling_params}
\bgroup
\def\arraystretch{1.25}
\begin{tabular}{lcccc}
\hline
\multicolumn{1}{c}{System} & $\sigma$ & uniform sampling (\%) & & data used (\%) \\ \hline
Benzene & 90 & 50 & & 0.10 \\
Al & 40 & 60 & & 0.50 \\
Mo & 30 & 40 & & 0.12 \\
$1T^\prime$-MoS$_2$ & 40 & 60 & & 0.05 \\ \hline
\end{tabular}
\egroup
\end{table}

\FloatBarrier
\section*{DATA AVAILABILITY}
The data used to train and test the models (DFT charge density, structure files, and trained models) is available via Zenodo \cite{zenodo}.

\section*{CODE AVAILABILITY}
Scripts and related code for calculating the Jacobi-Legendre grid-based linear expansion are available at \url{https://github.com/StefanoSanvitoGroup/MLdensity}.

\begin{acknowledgments}
This work was supported by S\~ao Paulo Research Foundation (FAPESP) (Grants no. 2021/12204-6, 2019/04527-0, and 2017/02317-2), and by the Irish Research Council Advanced Laureate Award (IRCLA/2019/127). We acknowledge the DJEI/DES/SFI/HEA Irish Centre for High-End Computing (ICHEC) and Trinity Centre for High Performance Computing (TCHPC) for the provision of computational resources. We acknowledge support from ICHEC via the academic flagship program (Project Number - EuroCC-AF-3). We acknowledge NVIDIA Academic Hardware Grant Program for providing graphics processing units. 
\end{acknowledgments}

\section*{AUTHOR CONTRIBUTIONS}
S.S. conceived the idea of a machine learning model for the charge density as starting guess for DFT calculations. M.D. developed the Jacobi-Legendre representation of the grid points and the many-body linear expansion of the charge density. B.F. and U.P. implemented the grid-based JL representation and linear model. B.F. performed all DFT calculations, ML training, and testing and implemented related code for results analysis. S.S. and A.F. supervised the work. All authors contributed to discussions, writing, and revision of the manuscript.

\section*{Competing Interests}
The Authors declare no Competing Financial or Non-Financial Interests.

\clearpage
\includepdf[pages=1]{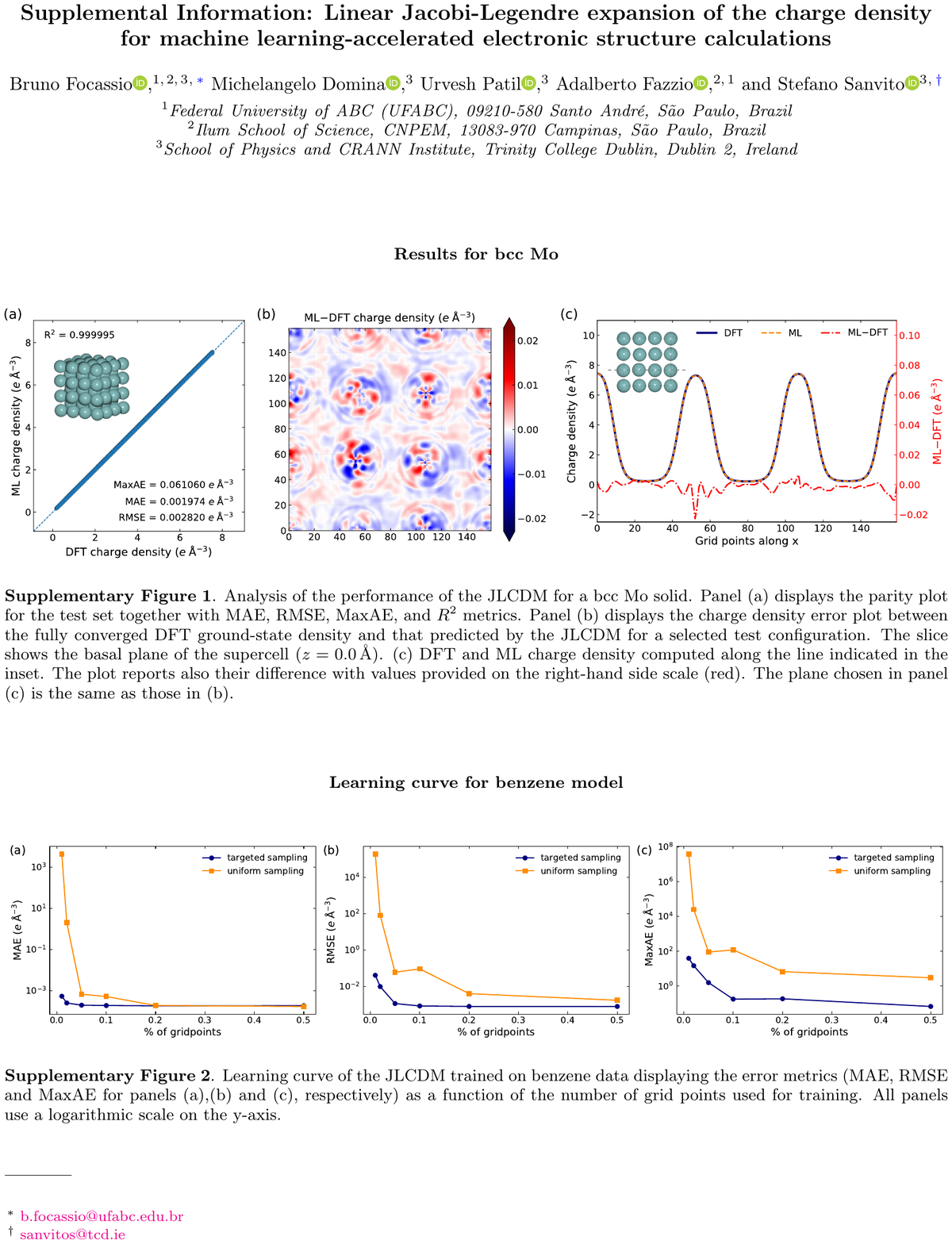}
\clearpage
\includepdf[pages=2]{41524_2023_1053_MOESM1_ESM.pdf}

\end{document}